\documentclass[aps,prl,reprint, amsmath, amsthm, amssymb, superscriptaddress]{revtex4-2}
\usepackage{float}
\usepackage{color}
\usepackage[group-separator={,},group-minimum-digits={3}]{siunitx}
\definecolor{Red}{rgb}{1,0,0}
\definecolor{Blue}{rgb}{0,0,1}
\definecolor{Olive}{rgb}{0.41,0.55,0.13}
\definecolor{Yarok}{rgb}{0,0.5,0}
\definecolor{Green}{rgb}{0,1,0}
\definecolor{MGreen}{rgb}{0,0.8,0}
\definecolor{DGreen}{rgb}{0,0.55,0}
\definecolor{Yellow}{rgb}{1,1,0}
\definecolor{Cyan}{rgb}{0,1,1}
\definecolor{Magenta}{rgb}{1,0,1}
\definecolor{Orange}{rgb}{1,.5,0}
\definecolor{Violet}{rgb}{.5,0,.5}
\definecolor{Purple}{rgb}{.75,0,.25}
\definecolor{Brown}{rgb}{.75,.5,.25}
\definecolor{Grey}{rgb}{.7,.7,.7}
\definecolor{Black}{rgb}{0,0,0}

\usepackage{graphicx}
\begin{document}


\title{Universality classes for percolation models with long-range correlations}

\author{Christopher Chalhoub}
\affiliation{Imperial College London, Department of Mathematics, London SW7 2AZ, United Kingdom}
\email{christopher.chalhoub21@imperial.ac.uk p.rodriguez@imperial.ac.uk} 

\author{Alexander Drewitz}
\affiliation{Universit\"at zu K\"oln, Department Mathematik/Informatik, 50931 K\"oln, Germany}\email[]{adrewitz@uni-koeln.de}
\affiliation{NYU Shanghai, NYU-ECNU Institute of Mathematical Sciences,  China}

\author{Alexis Pr\'evost}
\affiliation{University of Geneva, Section of Mathematics, 
1211 Geneva, Switzerland}
\email{alexis.prevost@unige.ch}

\author{Pierre-Fran\c cois Rodriguez}
\affiliation{Imperial College London, Department of Mathematics, London SW7 2AZ, United Kingdom}
\email{p.rodriguez@imperial.ac.uk} 


\date{\today}

\begin{abstract}
We consider a class of percolation models where the local occupation variables have long-range correlations decaying as a power law $\sim r^{-a}$ at large distances $r$, for some $0< a< d$ where $d$ is the underlying spatial dimension. For several of these models, we present both, rigorous analytical results and matching simulations that determine the critical exponents characterizing the fixed point associated to their phase transition, which is of second order. The exact values we obtain are rational functions of the two parameters $a$ and $d$ alone, and do not depend on the specifics of the model.
\end{abstract}

\maketitle

\begin{table*}
\caption{\label{tab:exponents} Critical exponents as a function of the roughness parameter $a \,(\leq d-2)$ and the dimension $d$.}
\begin{ruledtabular}
\begin{tabular}{ccl|cccccccccc}
&Exponent & & $\alpha$ & $\beta$\footnotemark[1] 
 & $\gamma$ & $\delta$ & $\Delta$  & $\rho$ & $\nu$ & $\eta$\footnotemark[1]  & $\kappa$\footnotemark[1]$^,$\footnotemark[2] &\\ 
\hline
&Value & & $2-\frac{2d}a$ & 1 & $\frac{2(d-a)}a$ & $\frac{2d}a-1$&$\frac{2d}a-1$&$\frac2a$&$\frac2a$&$a-d+2$ &$\frac12$ &\\
\end{tabular}
\end{ruledtabular}
\footnotetext[1]{Exponents without this superscript hold for small enough $a$, in particular for $a\leq \frac{d}{2}$.}
\footnotetext[2]{The capacity exponent $\kappa$ is introduced below \eqref{eq:density}.}
\end{table*}

\paragraph{Introduction.--}Due to the fundamental role that phase transitions and spontaneous symmetry breaking play in physics, a significant amount of work has been devoted to the analysis of these phenomena near their critical points, both by exact as well as computational methods~\footnote{These include (among others): renormalization group methods, see \cite{Pelissetto:2000ek} for a review, see also \cite{zbMATH07335095}; CFT techniques, notably in three dimensions, see the review \cite{Poland:2018epd} and refs.~therein; exactly solvable models \cite{baxter-book}; recent rigorous progress on conformal invariance and universality in two dimensions \cite{SW01,*lsw-ust,*zbMATH05808591,*zbMATH01506590,*zbMATH06084027}.}. In this Letter we present both rigorous and numerical results that unveil the universality classes of percolation models in the presence of long-range correlations \cite{RevModPhys.64.961,AGK14,SABERI20151}. Our results determine the full set of associated critical exponents, which are explicit algebraic functions of two parameters alone, the spatial dimension $d$ of the system and the roughness exponent $a$ governing the strength of interactions, see Table~\ref{tab:exponents} for a summary. These values rigorously confirm various classical predictions from the physics literature, notably a  criterion by Weinrib-Halperin \cite{PhysRevB.27.413,*PhysRevB.29.387} that forecasts the value of the correlation length exponent $\nu$ for such systems; see also \cite{Chayes:1986ju,zbMATH04076279} for related (partial) rigorous results in the short-range case. 
As further discussed below, the values we obtain are in accordance with those of the spherical model derived by Joyce \cite{PhysRev.146.349}, and thus also correspond to the heuristics by Fisher-Ma-Nickel \cite{PhysRevLett.29.917} for $n$-component spin systems. Furthermore, they certify a number of numerical results (see below for a full list), including  for instance simulations by Abete-de Candia-Lairez-Coniglio \cite{PhysRevLett.93.228301}, see also the recent conjecture (20) in \cite{PhysRevE.108.024312,*[{ see also }]FG24} by Feshanjerdi-Masoudi-Grassberger-Ebrahimi.

Our rigorous analysis is driven by the identification of a specific model, the M-GFF, which enjoys certain integrability properties and is amenable to a rigorous study.  Subsequent simulations for other models with different microscopic occupation rules but similar correlation structure provide us with critical exponents that coincide with our rigorous findings for M-GFF. This strongly suggests that our rigorous results also apply to these other models, and hence represents a substantive indicator of universality.

\paragraph{The models.--}We now introduce three models which  exhibit a percolation phase transition driven by a `temperature' parameter $u.$
In analogy with ferromagnetic (e.g.~Ising-type) spin systems, the  onset of a magnetized phase across the Curie temperature corresponds in the present context to the emergence of an infinite cluster when $u$ crosses a critical threshold $u_*$. 

The three models we consider are all defined on the sites of an infinite regular lattice of dimension $d>2$ in terms of occupation variables $\Psi_u(x)$ with values in $\{0,1\} = \{\text{closed}, \text{open}\}$ that depend monotonically on the
parameter $u$ which regulates the local density $ \text{Prob}(\Psi_u(x)=1).$ 
We write $\mathcal{C}_u$ for the open cluster of the origin in the resulting percolation configuration ($\mathcal{C}_u$ may well be empty in case the origin is closed, i.e.~$\Psi_u(0)=0$). Percolation studies whether or not $\mathcal{C}_u$ is infinite or not. The corresponding order parameter is 
\begin{equation}
\label{eq:theta}
\theta(u)= \text{Prob}(\mathcal{C}_u \text{ is infinite})
\end{equation}
(it is the analogue of the average magnetization for spin systems).
Our models will be defined in such a way that $\theta(u)$ is decreasing in $u$. The associated critical point is therefore
\begin{equation}
\label{eq:u_*}
u_*=\inf \{u : \theta(u)=0\}.
\end{equation}
In words, \eqref{eq:u_*} implies that for $u>u_*$ (the sub-critical/disordered phase), the cluster $\mathcal{C}_u$ of the origin is finite with probability one. On the contrary for $u<u_*$ (the super-critical/ordered phase), $\mathcal{C}_u$ is infinite with positive probability; equivalently, an infinite cluster is present somewhere on the lattice (not necessarily at the origin) with probability one.

 The microscopic descriptions of the models provided below will all lead to an asymptotic long-range (LR) decay of the correlation function of the occupation variables having the form
\begin{equation}
\label{eq:LR}
\tag{$\text{LR}_a$}
\langle \Psi_u(x) \Psi_u(y) \rangle -\langle \Psi_u(x) \rangle \langle \Psi_u(y) \rangle \sim |x-y|^{-a},
\end{equation}
for some exponent $a$ satisfying $0<a < d$. The correlations \eqref{eq:LR} are present for $\textit{all}$ values of the parameter $u$ and not specific to the (near-)critical regime $u \approx u_*$. Their presence characterizes the long-range class $(\text{LR}_a)$.\\

\textbf{Model 1: Vacant set of random walk (RW).} This model was introduced in \cite{PhysRevLett.93.228301} to study enzyme gel degradation. It is best explained in its finite-volume version, on a $d$-dimensional torus $\mathbb{T}_N$ of side length $N \gg 1.$ Consider $X=(X_n)_{n \geq 0}$ the random walk with usual nearest-neighbor hopping on $\mathbb{T}_N$, started from a uniformly chosen point. The walk evolves until time $t_u= uN^d$ and perforates the lattice, i.e.~one sets
\begin{equation}
\label{eq:RW}
\Psi_u(x)=\begin{cases}
1, & \text{if $x$ is not visited by $X$ until time $t_u$,}\\
0, & \text{otherwise.}
\end{cases}
\end{equation}
Thus $\mathcal{C}_u$ corresponds to the cluster of the origin in the vacant set of the walk. As $N \to \infty$, this model has a local  limit which is defined on the infinite lattice, the vacant set of random interlacements \cite{MR2680403}. The threshold $u_*$ in \eqref{eq:u_*} is defined as the transition point in this infinite model, and corresponds  \cite{MR2386070,*MR2838338,*MR3563197} to the emergence of a giant component in $\mathcal{C}_u$ on $\mathbb{T}_N$ that scales linearly with the system size.\\

\textbf{Model 2: Gaussian free field (GFF).} One considers the massless free field $\phi$ on the  Euclidean lattice, that is, the mean zero scalar Gaussian field $\phi=(\phi(x))$ with covariance
$\langle \phi(x)\phi(y) \rangle= (-\Delta)^{-1}(x,y)$, where $\Delta$ denotes the lattice Laplacian. 
Flooding the landscape $\phi$ from below up to height $u$  induces a natural percolation problem for its remaining dry parts,
i.e.~its excursion sets above height $u \in \mathbb{R}$, which corresponds to setting
\begin{equation}
\label{eq:GFF}
\Psi_u(x)=\begin{cases}
1, & \text{if $\phi(x) \geq u$,}\\
0, & \text{otherwise.}
\end{cases}
\end{equation}
The occupation variables in \eqref{eq:GFF} inherit the LR-dependence from $\phi$. The study of this model was initiated in \cite{LEBOWITZ1986194, *MR914444}, and more recently re-instigated in \cite{MR3053773}; it has since then received considerable attention, in particular in the mathematics community, see \cite{MR3325312,*MR3339867,*MR3417515,*Sz-16,*10.1214/16-AIHP799,*sznitman2018macroscopic,*DrePreRod2,*nitzschner2018,*chiarini2018entropic,*DCGRS20,*10.1214/20-EJP532}. \\

\textbf{Model 3: Metric graph GFF (M-GFF).} Model~3 is a variation of Model 2 introduced in \cite{MR3502602}. Albeit slightly harder to define, it enjoys greater integrability properties, as will become clear in the next section. The M-GFF is naturally defined as a \textit{bond} percolation model. The starting point is still the massless free field $\phi$, but one now superimposes the following (quenched)  \textit{bond} disorder: given $\phi$, each bond $e=\{x,y\}$ of the lattice is declared open independently with probability
\begin{equation}
\label{eq:def}
1-\exp\big\{-(\phi(x)-u)_+(\phi(y)-u)_+ \big\},
\end{equation}
where again $u$ is interpreted as a varying real height parameter and $v_+=\max\{ v,0\}$. Thus, $e$ can only possibly be opened if $\phi$ exceeds the value $u$ at both its endpoints, and then it does so with probability given by \eqref{eq:def}. Incidentally, the additional disorder \eqref{eq:def} actually corresponds to replacing the bond $\{x,y\}$ by a continuous line segment of length $1$, and asking that a Brownian bridge on this line segment with values $\phi(x)$ and $\phi(y)$ at the endpoints always stays above a horizontal barrier placed at height $u$. The cluster $\mathcal{C}_u$ is then defined as the set of vertices connected to the origin by open edges.\\

\paragraph{Rigorous results for M-GFF.--}The M-GFF allows to determine various critical exponents rigorously below the upper critical dimension, which is remarkable given that it is a non-planar model. We return to this below, and start by summarizing some of the results in the following theorem; we refer to~\cite{DPR21,*[{ see also }]DrePreRod3,*Pre1,*DrePreRod6,MR3502602} for full accompanying mathematical results and proofs. Below, the truncated two-point function $\tau_u^{\textrm{tr}}(x,y)$ refers to the probability that $x$ and $y$ are connected by a finite open cluster. The following results hold in fact on any graph with volume growth of balls of the form $B_r \sim r^d$, for (not necessarily integer-valued) dimension $d>2$. Regarding the following result, we observe that for any such $d$, any $a$ with $0 < a \leq d-2$ can actually be realized by  choosing the underlying graph adequately \cite{MR2076770}.\\
 
\textit{Theorem}-- For M-GFF, if $\phi$ is of class \eqref{eq:LR} for some $0 < a \leq \frac{d}{2}$, one has:
\\
(a) For all $u$ and all points $x,y$,
\begin{equation}\label{eq:2point}
\tau_u^{\textrm{tr}}(x,y) =  |x-y|^{-a} \, G_a\big(|x-y|/|u-u_*|^{-\frac2a}\big),
\end{equation}
where
$$
-\log(G_a(t))\propto\begin{cases}
t, & \text{ if } a> 1,\\
\frac{t}{\log(1+t)}, & \text{ if } a= 1,\\
t^a, & \text{ if } a< 1.
\end{cases}
$$
(b) For all $r \geq 1$, with $\psi(r)$ denoting the probability that a point is connected to distance $r$ at the critical point $u=u_*$,
\begin{equation}\label{eq:one-arm}
\psi(r) \propto r^{-\frac{a}{2}} \text{ as } r \to \infty,
\end{equation}
and $\psi$ exhibits the same off-critical scaling as in (a).\\

The above Theorem entails key information on the critical exponents associated to the M-GFF. 
From the bounds derived in (a), the correlation length of the model can be read off as $\xi= |u-u_*|^{-\frac2{a}}$, and this determines the corresponding exponent $\nu \stackrel{\text{def.}}{=} - \lim_{u \to u_* } \frac{\log \xi}{\log|u-u_*|} = \frac2a$. These findings are in line with the extension of the Harris criterion to the class \eqref{eq:LR} advocated by Weinrib-Halperin in \cite{PhysRevB.27.413,*PhysRevB.29.387} 
and establish its rigorous justification. Similarly, at criticality, (b)
yields the critical exponent $ \frac{1}{\rho} \stackrel{\text{def.}}{=}- \lim_{r\to \infty} \frac{\log \psi(r)}{\log r} =\frac a2$.  Moreover, by integrating \eqref{eq:2point} over $y$, one further deduces for $a<d/2$ that the limit 
\[
 \gamma \stackrel{\textnormal{def.}}{=}- \lim_{u \to u_*} \frac{\log( \langle |\mathcal C_u|1\{|\mathcal{C}_u|<\infty\} \rangle )}{\log |u-u_*|} \ \big(= \frac{2d}a-2 \big)
 \] 
exists (some additional log corrections when $a=d/2$ prevent to draw this conclusion in this case). In particular, it follows that Fisher's relation $
\gamma= \nu(2-\eta)$ is satisfied,
where $\eta$ describes the deviation of the two-point function exponent from the exponent of  Green's function decay. We emphasize that such scaling relations, let alone the values of the exponent they relate, are notoriously hard to establish rigorously. 

The proof of the Theorem exhibits an integrable 
quantity
for the M-GFF, the electrostatic capacity of the cluster $\mathcal{C}_u$, which is denoted by $\text{cap}(\mathcal{C}_u)$ in the sequel. This quantity is defined through the variational principle  
\begin{equation*}
\text{cap}(\mathcal{C}_u) = \Big(\inf_\mu \sum_{x,y \in \mathcal{C}_u} (-\Delta)^{-1}(x,y)\mu(x)\mu(y)\Big)^{-1},
\end{equation*}
where the infimum ranges over probability measures $\mu$ on $\mathcal{C}_u$ and $\Delta$ denotes the lattice Laplacian.
The particular role of this observable originates in the deep connections linking $\phi$ and potential theory through the energy functional $\sum_{e=\{x,y\}} |\nabla\phi(e)|^2$ defining the Boltzmann weight.
The observable $\text{cap}(\mathcal{C}_u)$ is integrable in the sense that its probability density $f$ has an explicit form and  satisfies
\begin{equation}
\label{eq:density}
f(t) \propto  \ t^{-3/2} e^{-(u-u_*)^2 t/2}, \text{ as } t \to \infty.
\end{equation}
 In particular, the critical capacity satisfies $\text{Prob}(\mathrm{cap}(\mathcal{C}_{u_*})\geq t)\propto t^{-\kappa}$ with $\kappa=1/2.$ Furthermore, the explicit formula for $f$ derived in \cite{DrePreRod3}
yields that
$\theta(u)= \text{Prob}(\text{cap}(\mathcal{C}_u)=\infty) \propto |u- u_*|^{\beta}$ for $u \uparrow u_*$ with $\beta=1$.
One can now feed the values of exponents into the usual scaling and hyperscaling relations, which leads to a vastly overdetermined system. Yet, this system has a unique solution, summarized in Table~\ref{tab:exponents}. If the random walk is diffusive---that is $a=d-2$---then $a$ can be eliminated, and the values in Table~\ref{tab:exponents} converge towards the expected mean-field values as $d \uparrow 6$. It was recently proved \cite{cai2023onearm,*[{see also }]Werner2021,*GangulyNam} that \eqref{eq:one-arm} does in fact saturate in high dimensions, in that the mean-field scaling $\psi(r)\propto r^{-2}$ holds for all $d> 6$.\\

\paragraph{Numerical results for RW and GFF.--}We now compare the above rigorous findings, which Table~\ref{tab:exponents} summarizes, with numerical results (both existing and new). We have run Monte-Carlo simulations for both RW and GFF on the torus $\mathbb{T}_N$ in $d=3$, on which $a=1$, for various length scales $N$ of the torus with $50 \leq N \leq 600$ and varying values of $u$ within the critical window. For $N \geq 250$,  we simulated \num[group-separator={,}]{6000} runs for each point; for $150 \leq N < 250$, there were \num[group-separator={,}]{15000} runs and for $N < 150$ we performed \num[group-separator={,}]{50000} runs. On the torus, the quantity $\theta(u)$ in \eqref{eq:theta} is replaced by $\theta_N(u)$,  obtained by modifying the condition that $\mathcal{C}_u$ is infinite to the requirement that $\mathcal{C}_u$ spans the torus, that is, its  projection on the first coordinate equals all of $\{1,\dots,N\}$. The simulations were used in combination with scaling arguments to compute the critical exponents $\beta$, $\gamma$ and $\nu$. We now present these findings.

\begin{figure}[t]
\includegraphics[scale=0.6]{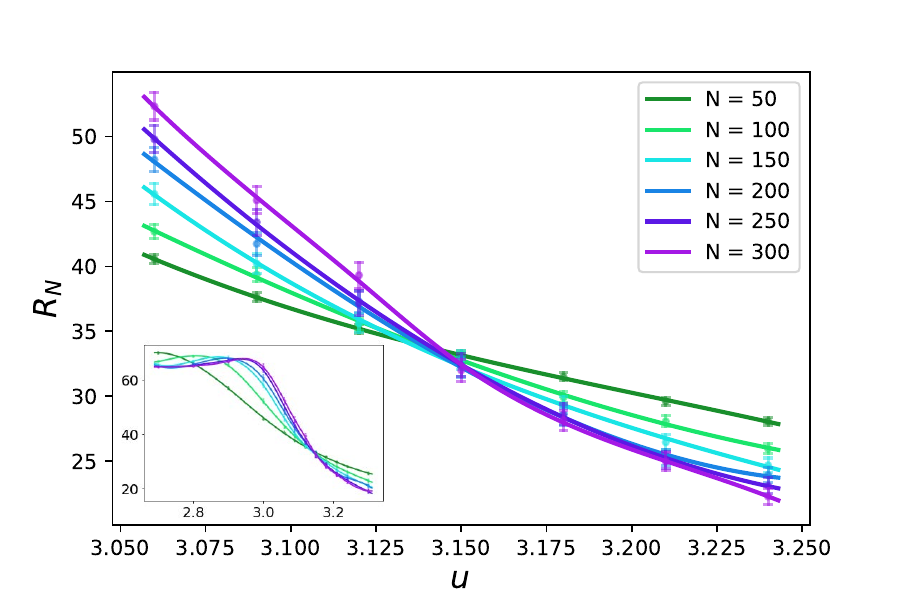}
\caption{\label{fig:R_N} $R_N$ vs.~$u$. The critical value $u=u_*$ corresponds to the point of intersection. For RW (depicted), one obtains $u_*=3.15\pm0.015$ and consequently $\frac{\gamma}{\nu}=2.05 \pm 0.07$.}
\end{figure}

\begin{figure}[b]
\includegraphics[scale=0.6]{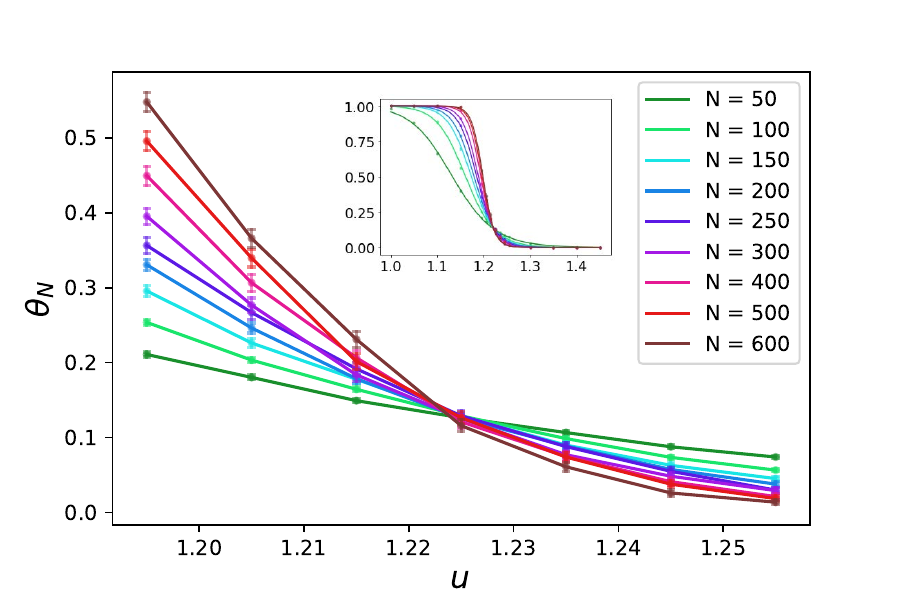}
\caption{\label{fig:inter} $\theta_N$ vs.\ $u$ for GFF. The intersection of the lines for different values of $N$ confirms the value $u_*=1.225 \pm 0.005$.}
\end{figure}

\begin{figure}[t]
\includegraphics[scale=0.6]{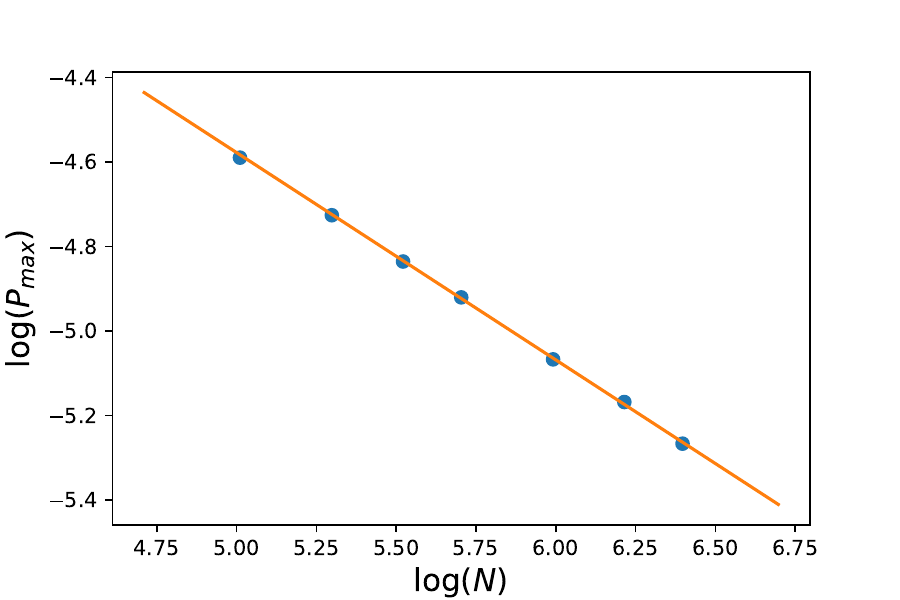}
\caption{\label{fig:P_max} $P_{\text{max}}(N)$, the density of the largest cluster at criticality, as a function of the box size $N$ for RW. By scaling $P_{\text{max}}(N)\sim \theta(u)\propto |u-u_*|^\beta$ where $u$ is s.t. $\xi(u)=|u-u_*|^{-1/\nu} =N$. Overall, this yields $\frac{\beta}{\nu}=0.49 \pm 0.07$.}
\end{figure}

Fig.~\ref{fig:R_N} shows the ratio $R_N=R_N(u)=\Gamma_{2N}(u)/\Gamma_N(u)$ as a function of $u$ for RW, where $\Gamma_N$ is the second moment of the cluster size distribution for $\mathbb{T}_{N}$. With a scaling ansatz, the parameter $u_*$ is inferred as the point at which $R_N$ is independent of $N$; see Fig.~\ref{fig:R_N}. This yields $u_*=3.15\pm 0.015$ for RW and $u_*=1.225\pm 0.005$ for GFF. Considering the intersection point of the curves mapping $u$ to $\theta_N(u)$ for different values $N$ further confirms this prediction, see Fig.~\ref{fig:inter} for GFF. The ratio $\frac{\gamma}{\nu}= \log_2 R_N(u_*) -d$ can then be directly computed and gives $\frac{\gamma}{\nu}=2.05 \pm 0.07$ for RW and $\frac{\gamma}{\nu}=1.98 \pm 0.04$ for GFF.
Let $P_{\text{max}}(N)=P_{\text{max}}(u_*,N)$ denote the empirical density of the largest cluster in $\mathbb{T}_N$ at $u_*$. Again by scaling one finds $-\frac{ \log P_{\text{max}}(N)}{\log N} = \frac{\beta}{\nu}$ yielding $\frac{\beta}{\nu}=0.49 \pm 0.07$ for RW and $\frac{\beta}{\nu}=0.51 \pm 0.05$ for GFF, see Fig~\ref{fig:P_max} for RW. 
\begin{figure}[t]
\includegraphics[scale=0.6]{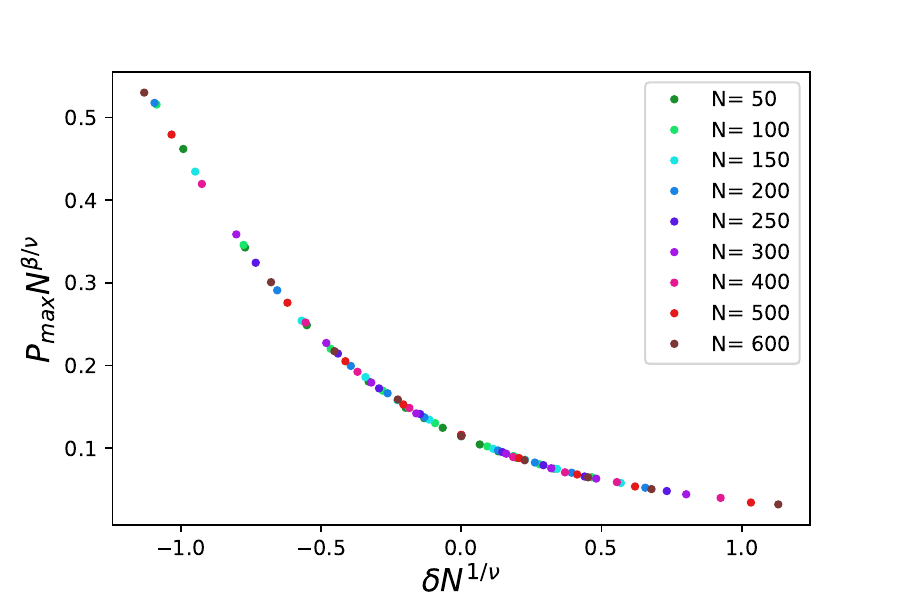}
\caption{\label{fig:collapse} Collapse for RW: $P_{\text{max}}(u, N)N^{\frac\beta\nu}$ as a function of $\delta N^\nu$, with $\delta=|u-u_*|$, becomes independent of $N$ for $\frac{\beta}{\nu}= 0.485\pm0.02$, $\nu=2.02\pm 0.08$.}
\end{figure}
An alternative method consists of plotting $P_{\text{max}}(u, N)N^{\frac\beta\nu}$ as a function of $\rho= \xi(u)N=|u-u_*|^{-\frac1{\nu}}N$ by which $P_{\text{max}}(u, N)N^{\frac\beta\nu}= G_a(\rho)$. One then fits the parameters such that the functions $G_a(\cdot)$ obtained for different values of $N$ collapse, i.e.~do not depend on $N$, see Fig~\ref{fig:collapse} for RW. This gives $\frac{\beta}{\nu}= 0.485\pm0.02$, $\nu=2.02\pm 0.08$ for RW and $\frac{\beta}{\nu}= 0.515\pm0.03$, $\nu=1.99\pm 0.07$ for GFF. 
A similar method applies to $(\text{Prob}(\Psi_u(0)=1)N^d)^{-1}\Gamma_N(u)$ as a function of $\rho$ and yields $\frac{\gamma}{\nu}= 2.05\pm0.07$, $\nu=1.95\pm0.11$, for RW and $\frac{\gamma}{\nu}= 2.00\pm0.08$, $\nu=1.94\pm0.1$ for GFF. All these simulated values are in accordance with those of Table~\ref{tab:exponents}.

The RW model has first been studied in \cite{PhysRevLett.93.228301}, where the numerical results $\nu=1.8\pm0.1$, $\beta=1.0\pm 0.1$ and $\gamma=3.4\pm 0.2$ were obtained. Although some of these values first seem far off from the ones from Table~\ref{tab:exponents} when $d=3$ and $a=1$, they are in fact in accordance with the simulated value of $a=1.15\pm 0.05$ from \cite{PhysRevLett.93.228301}, which is probably due to the small size $N=60$ of the lattice therein. More recent simulations  \cite{PhysRevE.100.022125} find values of $\nu$ in accordance with \cite{PhysRevB.27.413,*PhysRevB.29.387} when $d\in{\{3,4,5\}}$, which corresponds to $a\in{\{1,2,3\}}$; see also the numerics announced in \cite{PhysRevE.108.024312} (Ref.~[49] therein) in support of their conjecture (20), which matches Table~\ref{tab:exponents}.

Some simulations for the GFF model have also been conducted in \cite{PhysRevE.74.031120}, and they also find values of $\nu$ in accordance with \cite{PhysRevB.27.413,*PhysRevB.29.387}. However the value of $\beta/\nu=0.60\pm 0.01$ and $\gamma/\nu=1.8\pm 0.1$ when $a=1$ obtained therein seem to differ from the one indicated by both Table~\ref{tab:exponents} and our simulations. It would however be rather surprising that $\beta$ is larger than its mean-field value (for instance the curve $\theta(u)$ for the order parameter in \eqref{eq:theta} would fail to be convex); this partially motivated the new simulations for GFF in the current work. We also refer to \cite{GRS20,*[{see also }][ for the RW model.]prevost2023passage} for rigorous results which suggest that $\nu=2$ in dimension three. Finally in \cite{PhysRevE.108.024312} a RW model slightly different from Model 1 was studied, where the RW moves at random using the same moves as a knight in chess (in particular, its trajectory is not necessarily connected), and one studies percolation for the set of points visited by the random walk, instead of the vacant set. In dimension three, the  values $\beta/\nu=0.498\pm 0.005$ and $\nu=1.99\pm 0.01$ are obtained numerically and the authors conjecture that the exact values for $\beta$ and $\nu$ are in fact the ones from Table~\ref{tab:exponents}. We also refer to \cite{MuiSev} for results on general Gaussian fields which suggest that $\nu=2/a$ for $a\leq 1$.

\paragraph{Discussion.--}As explained in \cite{PhysRevB.27.413,*PhysRevB.29.387}, see also \cite{PhysRevE.96.062125} for some simulations, it is expected that the equality $\nu=\frac2a$ holds until $\nu$ reaches its short range value and then remains constant. This sticking phenomenon is reminiscent of the one for $\eta$ for $n$-component spin systems of LR-type, as first observed by Sak \cite{PhysRevB.8.281,*[{ see also }]MR3269693,*MR3723429}, see also Fisher et al.\ \cite{PhysRevLett.29.917}. Their predicted value $\eta=a-d+2$ (for long-range interactions decaying as $r^{-2d+a}$, so that the associated GFF satisfies \eqref{eq:LR}) matches the value of Table~\ref{tab:exponents}. Feeding it alone into the \emph{hyper}scaling relation $(2-\eta)(\delta+1)= d(\delta-1)$ yields
the value of $\delta = \frac{2d}{a}-1$ from Table~\ref{tab:exponents}, and we refer to \cite{MR4341081,*hutchcroft2021critical,*hutchcroft2022critical} for similar results in the context of independent long-range percolation. Letting $n\rightarrow\infty$, one recovers the spherical model \cite{PhysRev.176.718,*KacThompson,*10.1143/PTP.49.424,*MR3772040}, whose exponents \cite{PhysRev.86.821,*baxter-book,*PhysRev.146.349} all coincide with the ones from Table~\ref{tab:exponents} for a certain regime of parameters (up to the usual rescaling of the exponents when passing from spin systems to percolation). 
It would be interesting to understand whether there are deeper links between these models and the M-GFF, as well as whether all long-range models share a similar behavior when $d >2$; see also \cite{PhysRevE.108.024312,  PhysRevE.88.052102} for related results when $d=2$.

\paragraph{Conclusion--}
Our main contribution is the combination of our rigorous results for M-GFF  with numerical simulations for RW and GFF. This strongly suggests that all these models lie in the same universality class and that the exact critical exponents of M-GFF from Table \ref{tab:exponents} also apply to the other models. With the help of the integrable observable of the cluster capacity, we rigorously derived that M-GFF for $0<a \leq\min(\frac{d}{2},(d-2))$ undergoes a continuous phase transition with critical exponents summarized in Table~\ref{tab:exponents}. Moreover, we showed numerically that both the RW and GFF model on the cubic lattice have the same exponents $\beta$, $\gamma$ and $\nu$ as  M-GFF. 
Taking advantage of scaling and hyperscaling, we inferred that they are in the same universality class, which essentially coincides in the long-range regime with the universality class of the spherical model. Recent progress \cite{PhysRevE.108.024312} provides numerical evidence for further models satisfying \eqref{eq:LR} with small enough $a$ to belong to this universality class. 
Intriguing challenges in understanding this universality class remain open, numerically as well as rigorously. One very interesting question is to improve our understanding of the crossover to short range universality classes. That is, to assess how far the above set of exponents describes the truth for larger values of $a$, which depends on the underlying lattice through the values of its associated short-range exponents. One case in point is to show that the exponents for any of Models 1-3 on the hypercubic lattice in  $d=5$ are described by the values in Table~\ref{tab:exponents} for $a=d-2=3$.

\medskip

\paragraph{Acknowledgments--} The authors would like to thank Sebastian Diehl and Joachim Krug for valuable comments on a first draft of the paper. The research of AD has been supported by the Deutsche Forschungsgemeinschaft (DFG) grant DR 1096/2-1. AP was supported by the Swiss NSF. CC was supported by the EPSRC Centre for Doctoral Training in Mathematics of Random Systems: Analysis, Modelling and Simulation (EP/S023925/1).

\bibliography{critexp}

\end{document}